\documentclass[10pt,draftcls,onecolumn]{IEEEtran}
\usepackage{cite}
\usepackage{amsmath,graphicx,amssymb}
\usepackage{MnSymbol}
\usepackage{wasysym}

\newtheorem{thm}{Theorem}
\newtheorem{defn}{Definition}

\makeatletter
\newcommand{\rom}[1]{\lowercase\expandafter{\romannumeral #1\relax}}
\makeatother

\begin{document}
\title{Analysis of Degrees of Freedom of Wideband Random Multipath Fields Observed Over Time and Space Windows}
\author{\authorblockN{Farhana Bashar, S.M. Akramus Salehin and Thushara D. Abhayapala }\\
\authorblockA{Research School of Engineering, \\
The Australian National University (ANU) \\
Canberra 0200 ACT, Australia}}

\maketitle
%
%
%

%
\begin{abstract}
In multipath systems, available degrees of freedom can be considered
as a key performance indicator, since the channel capacity grows
linearly with the available degrees of freedom. However, a
fundamental question arises: given a size limitation on the
observable region, what is the intrinsic number of degrees of
freedom available in a wideband random multipath wavefield observed
over a finite time interval? In this paper, we focus on answering
this question by modelling the wavefield as a sum of orthogonal
waveforms or spatial orders. We show that for each spatial order,
(i) the observable wavefield is band limited within an effective
bandwidth rather than the given bandwidth and (ii) the observation
time varies from the given observation time. These findings show the
strong coupling between space and time as well as space and
bandwidth. In effect, for spatially diverse multipath wavefields,
the classical degrees of freedom result of {\em time-bandwidth}
product does not directly extend to {\em time-space-bandwidth}
product.
\end{abstract}
\section{Introduction}

In multipath wireless communication systems, the use of the spatial
aspects of multipaths can ensure improved system performances
\cite{Foschini}. The study of the spatial aspects of
multipath fields, thus, becomes an important thread of research in
wireless communications and signal processing, and has more recently
been addressed by
\cite{rodTDjones,PoonBrodersenTse,FranceschettiChakraborty09,AyferOlivierTse12,Franceschetti2012}.
None of these approaches, however, provide clear view of the
interrelationships between space, frequency and time that affects
the degrees of freedom of multipath fields.

In this work, we study a band limited random multipath field
observed over a limited source-free region of space over a finite time
window. The observable multipath wavefield is considered to be
farfield, and we study this from a physical wavefield perspective.
In particular, the underlying physics of free space propagation is
used to model the multipath field as a sum of orthogonal waveforms
or spatial orders. This mathematical framework is similarly used in
\cite{hanlenTD}. However, in comparison, our derived result is more
accurate, since we have considered the effect of available spatial
information on the observation time. Moreover, the results provided
in \cite{hanlenTD}, are derived by using a geometrical argument to
extend the narrowband degrees of freedom result of \cite{rodTDjones}
to a broadband scenario and resulted in a complicated formula.
Further, it is unclear, how the usable (effective) bandwidth varies
from the given frequency bandwidth for the different spatial orders. In
this work, on the contrary, the degrees of freedom of wideband
multipath fields is evaluated in a simple manner, and we derive that
the wavefield is bandlimited within an effective frequency bandwidth
for each spatial order.

The work \cite{FranceschettiChakraborty09} characterized
multi-antenna systems in a wideband transmission regime, but how the
coupling between space and time as well as space and frequency
affects the information content of the waves was left as an open and
important problem. We show that the effective frequency bandwidth of
each spatial order is essentially related to the spatial dimension
of the observable field and varies from the frequency bandwidth of
the channel. Our results indicate that even though for lower spatial
orders, effective bandwidth is equal to the given frequency
bandwidth, for higher orders, effective bandwidth is less than the
given frequency bandwidth. Moreover, we show that the effective
observation time is independent of spatial order and is related to
the finite size of the observation region. These findings clearly
indicate the strong link between space and time as well as space and
frequency in spatially diverse multipath fields. These findings also
indicate that the classical degrees of freedom result of {\em
time-bandwidth} product does not directly extend to {\em
time-space-bandwidth} product as shown in \cite{PoonBrodersenTse,
Franceschetti2012}, rather the degrees of freedom of any particular
spatial order $n$ can be expressed as a product of {\em effective
observation time} and {\em effective bandwidth of the $n^{th}$
order}.  If we denote the degrees of freedom of the $n^{th}$ order
as $D_n$, the total degrees of freedom can be calculated by $\sum_n
D_n$, for all possible values of $n$.

In recent times, the study of degrees of freedom of distributed MIMO
communications (e.g., \cite{AyferOlivierTse12}) has gained a lot of
attention. In distributed MIMO systems, the users cooperate in
clusters to receive information. In such scenarios, our derived
results indicate that the lower spatial orders (independent
channels) can utilize the full bandwidth, whereas, the higher orders
can utilize only a fraction of the given bandwidth. We envisage that
our results will be useful in characterizing the degrees of freedom
of distributed or large scale MIMO systems.

\section{Random $2D$ Multipath Fields}
\label{waveform}

We consider a wireless multipath wavefield band limited to
$[F_0-W,F_0+W]$ and observed within a $2D$ disk region of radius $R$
over a finite time interval $[0,T]$. The observable multipath
wavefield within this channel is assumed to be farfield and is
generated by a source or distribution of sources and scatterers that
exist outside the region of interest.

\subsection{Multipath Plane Wave Representation}

Let $\Psi(\textbf{\emph{x}},\omega)$ denote a finite complex-valued
wideband multipath field in the region of interest where
$\textit{\textbf{x}}\equiv (r, \phi_x)$ represents a position vector
within the $2D$ observation region, $r=\| \textit{\textbf{x}}\| \leq
R$ denotes the Euclidean distance of $\textbf{\textit{x}}$ from the
origin, $\phi_x \in [0, 2\pi)$ is the azimuth angle of vector
$\textbf{\textit{x}}$ and $\omega$ is the angular frequency. Note
that a standard multipath model involves modeling every distinct
path explicitly as a plane wave. Hence, in this model, the multipath
field is generated by the superposition of plane waves as
\begin{eqnarray} \label{plane}
\Psi(\textbf{\emph{x}},\omega) = \int_{0}^{2\pi} a(\phi, \omega)
e^{ik \textbf{\emph{x}} \cdot \hat{\textbf{\emph{y}}}} d\phi
\end{eqnarray}
where $\hat{\textbf{\emph{y}}}\equiv (1, \phi)$, $k=\omega/c$ is the
scalar wavenumber, $c$ is the wave velocity and $a(\phi, \omega)$ is
the complex-valued gain of scatterers as a function of direction of
arrival $\phi \in [0, 2\pi)$ and angular frequency $\omega$.

\subsection{Orthogonal Basis Expansion}

We consider that the multipath field is generated by sources
external to the region of interest. Hence, we can use Jacobi-Anger
expansion \cite[p. 67]{coltonkress98} to represent the plane waves
in \eqref{plane} as

\begin{eqnarray} \label{orthogonal}
e^{ik \textbf{\emph{x}} \cdot \hat{\textbf{\emph{y}}}} =
\sum_{n=-\infty}^\infty i^n J_n(\frac{\omega}{c}r) e^{in
(\phi_x-\phi)}
\end{eqnarray}
where $J_n(\cdot)$ is the Bessel function of the first kind of
integer order $n$, and we can identify a countable set of orthogonal
basis functions over the $2D$ disk since
\begin{eqnarray}\label{e}
\int_{0}^{2\pi} \Phi_{n}(\textbf{\emph{x}})
\Phi^\ast_{m}(\textbf{\emph{x}}) d\phi_x=
\begin{cases} 2\pi, & \mbox{$n=m$} \\ 0, & \mbox{otherwise}
\end{cases}
\end{eqnarray}
where $\Phi_{n}(\textbf{\emph{x}})= e^{in \phi_x}$ and
$(\cdot)^\ast$ is the complex conjugate operator.
Observe that by
substituting \eqref{orthogonal} into \eqref{plane}, we obtain
\begin{eqnarray} \label{fieldi}
\Psi(\textbf{\emph{x}},\omega) = \sum_{n=-\infty}^\infty i^n
\alpha_n(\omega) J_n(\frac{\omega}{c}r) e^{in \phi_x}
\end{eqnarray}
where $\alpha_{n}(\omega)$ is the $n^{th}$ frequency dependent
coefficient and using \eqref{plane} can be defined as
\begin{eqnarray} \label{alpha}
\alpha_{n}(\omega) = \int_{0}^{2\pi} a(\phi, \omega) e^{-in \phi}
d\phi.
\end{eqnarray}

However, the information available about scatterers that generate
the multipath field $\Psi(\textbf{\emph{x}},\omega)$ is usually
limited. Thus, it is reasonable to represent the multipath field as
a random process. Referring to \eqref{plane}, the scattering gain
$a(\phi, \omega)$ is random and so is $\alpha_{n}(\omega)$ in
\eqref{alpha}. For mathematical simplicity of the analysis, we
assume uncorrelated scattering. As a result, the random gains
$a(\phi, \omega)$ and $a(\phi^\prime, \omega)$ at two distinct
incident angles and different frequencies are uncorrelated from each
other. Hence, using \eqref{alpha}, the uncorrelated scattering
assumption, and following a few intermediate steps, we find that
\begin{eqnarray} \label{alpha1}
E \{{|\alpha_{n}(\omega)|}^2\} = \int_{0}^{2\pi} E\{a(\phi, \omega)
a^\ast(\phi, \omega) \} d\phi
\end{eqnarray}
where $E\{\cdot\}$ represents the expectation operation.
Therefore, our observable field \eqref{fieldi} is a random multipath
field and can be represented by an infinite but countable set of
orthogonal basis functions.

\section{Observation Time of the Spatial Orders}
\label{effect1}
If the wavefield is generated by a single point/ source transmitting a time domain signal, then observing the resulting travelling wavefield over a time window $[0, T]$ within a $2D$ disk of radius $R$ captures information content of the time domain signal over a time interval $T+2R/c$. We formalize this statement for the $n^{th}$ order time domain signal $a_{n}(t)$ producing the $n^{th}$ order space-time wavefield $\psi_{n}(r, t)$ in the following theorem.
\begin{thm} [Observation time of the spatial orders] 
Given that the spatial orders $n$ are separated, observing a random
wireless multipath wavefield over a $2D$ disk of radius $R$ for a
time interval $T$ is equivalent to observing the information content of the underlying $n^{th}$ order time domain signal $a_{n}(t)$ over an effective
time interval
\begin{eqnarray}\label{TS}
T_{eff}=T+\frac{2R}{c}.
\end{eqnarray}
Further, this effective time interval $T_{eff}$ is not order
dependent and increases with the size of the observation region.
\end{thm}
\begin{IEEEproof} We can define the $n^{th}$ order
signal spectrum over space from \eqref{fieldi} as
\begin{eqnarray} \label{coefficienta}
\Psi_{n}(r,\omega) \triangleq \alpha_{n}(\omega)
J_n(\frac{\omega}{c}r).
\end{eqnarray}
Let $\psi_{n}(r, t)$ be the inverse Fourier transform of
$\Psi_{n}(r, \omega)$. Then, by taking the inverse Fourier transform
of \eqref{coefficienta}, we obtain
\begin{eqnarray}\label{OT}
\psi_{n}(r,t)= a_{n}(t)\ast U_{n}(\frac{tc}{r})
\end{eqnarray}
where the time domain coefficient $a_{n}(t)$ is the inverse
Fourier transform of $\alpha_{n}(\omega)$ which represents the $n^{th}$ order time domain signal and the Chebyshev
Polynomial of the first kind $U_{n}(tc/r)$ is the inverse Fourier
transform of $J_n(\omega r/c)$.

Observe that in \eqref{OT}, the $n^{th}$ order
signal over space $\psi_{n}(r,t)$ is a convolution between the $n^{th}$ order time domain signal $a_{n}(t)$ and the Chebyshev Polynomial $U_{n}(tc/r)$. Hence, any information content in the $n^{th}$ order signal over space $\psi_{n}(r,t)$ is contained in the $n^{th}$ order time domain signal $a_{n}(t)$.\\
We observe the $n^{th}$ order signal over space $\psi_{n}(r,t)$ over a time window $[0, T]$ within a $2D$ disk region of radius $R$. Moreover, the Chebyshev Polynomial $U_{n}(z)$ is defined only for $-1\leq z \leq 1$, as illustrated in Fig. \ref{LP}. Hence, $U_{n}(tc/r)$ is defined only for $-r/c\leq t \leq r/c$. As a result, if we consider that the $n^{th}$ order signal over space $\psi_{n}(r,t)$ is observed within a disk of radius $R$ over the time window $[0, T]$, it is possible to capture information about the $n^{th}$ order time domain signal $a_{n}(t)$ over the time window $[-R/c, T+R/c]$. This is equivalent to observing the $n^{th}$ order time domain signal $a_n(t)$ over a maximum time window of $T+2R/c$.
\begin{figure}[htb]
\centering
\includegraphics[width=.6\columnwidth, height= 5cm]{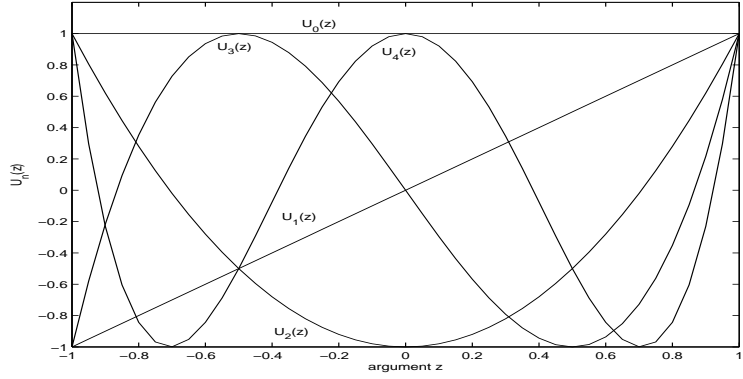}\\
\caption{Chebyshev Polynomial of the first kind $U_{n}(z)$ for
n=0,1,2,3,4.} \label{LP}
 \end{figure}
\end{IEEEproof}

\section{Effective Bandwidth of The $\lowercase{n^{th}}$ Order}
\label{effect2}

Ideally, if it is possible to measure signals with infinite
precision without noise, each spatial order $n$ would have an
effective bandwidth equal to the frequency bandwidth available,
i.e., from $F_0-W$ to $F_0+W$. However, in practical communication
systems, noise is present. As a result, it is not possible to detect
signals within the band of frequencies where the signal power to
noise ratio (SNR) drops below a certain threshold $\gamma$. This
threshold is dependent on the sensor sensitivity or the robustness
of the signal processing method to noise.

Lets consider $\eta_R(\phi_x, \omega)$ as the white Gaussian noise on
the circle (at radius $R$) associated with antenna/ sensor at an
angle $\phi_x$. Hence, the received signal on the circle is given by
\begin{eqnarray} \label{field}
\Psi(R, \phi_x,\omega) = \sum_{n=-\infty}^\infty i^n
\alpha_n(\omega) J_n(\frac{\omega}{c}R) e^{in \phi_x} +
\eta_R(\phi_x, \omega).
\end{eqnarray}
The following theorem proves that the white Gaussian noise power
remains the same at all frequencies in the modal expansion
\eqref{fieldi}.
\begin{thm} [White Gaussian Noise in $L^2$] Given a zero
mean white Gaussian noise with variance $\sigma_0 ^2$ in $L^2
(\mathbb{S}^1)$ represented by a random variable $\eta_R(\phi)$
where $\phi \in \mathbb{S}^1$, such that for any function
$\psi_i(\phi)\in L^2(\mathbb{S}^1)$ the complex scalar $\nu_i$
\begin{eqnarray*}
\nu_i \triangleq \int_{\mathbb{S}^1} \eta_R(\phi) \psi_i^\ast (\phi)
d\phi = \langle \eta_R(\phi), \psi_i(\phi)\rangle
\end{eqnarray*}
is also a zero mean Gaussian random variable with variance
$E\{{|\nu_i|}^2 \}= \sigma_0^2  \int_{\mathbb{S}^1}
{|\psi_i(\phi)|}^2 d\phi= \sigma_0^2 ({\|\psi_i(\phi)\|}_{L^2})^2$.
\cite[eqn 8.1.35]{gallager}
\end{thm}
\begin{defn} By taking $\psi_i(\phi)$
to be the orthogonal basis functions $e^{in \phi_x}$, the spatial
Fourier coefficients for the noise is
\begin{eqnarray} \label{noise}
\nu_{n}(\omega) = \int_{\mathbb{S}^1} \eta_R(\phi_x, \omega) e^{-in
\phi_x} d\phi_x
\end{eqnarray}
and  applying Theorem $2$, $\nu_{n}(\omega)$ are also zero mean
Gaussian random variables with variance $\sigma_0^2$.
\end{defn}

Based on Definition $1$, we can define the $n^{th}$ order received
signal at radius $R$ as
\begin{eqnarray} \label{coefficient}
\Psi_{n}(R,\omega) = \alpha_{n}(\omega) J_n(\frac{\omega}{c}R)+
\nu_{n}(\omega),
\end{eqnarray}
and we assume that the noise and the signal are independent of each
other.

\begin{figure}[htb]

\centering
\includegraphics[width=.6\columnwidth, height= 5cm]{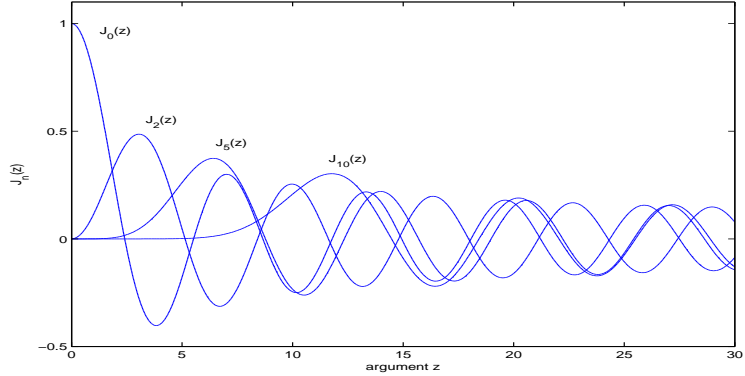}\\
\caption{Bessel functions of first kind $J_{n}(z)$ vs. argument $z$
for different values of $n$.} \label{bessel}
\end{figure}

Note that we can consider $\alpha_{n}(\cdot)$ as the $n^{th}$ order
signal spectrum that is defined only over the range $[F_0-W,
F_0+W]$. Also note that for a fixed value of the radius,
$J_n(\cdot)$ can be treated as a function of frequency. However, it
is evident from Fig. \ref{bessel} that except for the $0^{th}$
order, Bessel functions start small before increasing monotonically
to their maximum. Further, the Bessel functions start more slowly as
the order $n$ increases. Thus, for any particular order $|n|(>0)$
and for frequencies less than a critical frequency $F_n$, the
magnitude of the Bessel functions $|J_n(\cdot)|$ is negligible.

We note that our observable signal spectrum \eqref{coefficient} is a
product of $\alpha_{n}(\cdot)$ and $J_n(\cdot)$. Thus, for a fixed
value of radius, at each order $|n|>0$, the SNR is less than the
threshold $\gamma$ for frequencies less than a critical frequency
$F_n$. In effect, we can not detect the signal spectrum for
frequencies less than $F_n$. From Fig. \ref{bessel}, $J_0(\cdot)$ is
active within the frequency range $[0, \infty)$, hence, the
effective bandwidth of the $0^{th}$ order signal spectrum is $2W$.

\begin{thm}[Effective Bandwidth of the $n^{th}$ Order]
Given any wireless random multipath wavefield band limited to
$[F_0-W,F_0+W]$ and observed within a $2D$ disk region of radius
$R$, such that the wavefield is encoded in finite number of orders
$n< N_u$, the effective frequency bandwidth of the $n^{th}$ order signal
spectrum is given by
\begin{eqnarray}\label{BW}
W_n = \begin{cases} 2W, & \mbox{$n=0$} \\
F_0\negthickspace+\negthickspace W-\max\{F_0\negthickspace-\negthickspace
W, F_n\}, & \mbox{$|n|< N_u$} \\
0, & \mbox{\text{otherwise}} \end{cases}
\end{eqnarray}
where $N_u$ is the lowest order for which the critical frequency $F_n
> F_0+W$ and
\begin{equation} \label{cf}
F_n \geq \frac{nc}{e \pi R} + \frac{c}{2 e \pi R} \log \left(
\frac{\gamma}{{(SNR)}_{\max}} \right).
\end{equation}
with the threshold $\gamma$ depicting the ability of
the system to detect signals buried in noise and assuming that the power of the spectrum
$\alpha_{n}(\omega)$ is finite and bounded for all frequencies
$\omega$ and orders $n$, i.e., $E\{{|\alpha_{n}(\omega)|}^2\}\leq
P_{\max}$, the maximum SNR for any order $n$ is
\begin{eqnarray} \label{k}
{(SNR)}_{\max} = \frac{P_{\max}}{\sigma_0 ^2}.
\end{eqnarray}
\end{thm}

\begin{IEEEproof} From \eqref{fieldi} and \eqref{coefficient}, the observable random multipath field
can be represented by an infinite but countable set of orthogonal
basis functions as follows
\begin{eqnarray} \label{field1}
\Psi(\textbf{\emph{x}},\omega) = \sum_{n=-\infty}^\infty i^n
[\alpha_n(\omega) J_n(\frac{\omega}{c}R)+ \nu_{n}(\omega)] e^{in
\phi_x}
\end{eqnarray}
We  now define the average power of our observable multipath field
\eqref{field1} from all azimuth directions $\phi_x$ as

\begin{eqnarray}
\frac{1}{2 \pi}\negthickspace \int_{0}^{2\pi}\negthickspace
E\{{|\Psi(\textbf{\emph{x}},\omega)|}^2\} d\phi_x = \negthickspace
\sum_{n=-\infty}^\infty \negthickspace E
\{{|\alpha_{n}(\omega)|}^2\}
{|J_n(\frac{\omega}{c}R)|}^2\negthickspace+\negthickspace \sigma_0
^2.
\end{eqnarray}
%
The SNR at the $n^{th}$ order over the frequency band $[0,
\omega_n]$ with $\omega_n= 2\pi F_n$ is
\begin{eqnarray} \label{snr}
{(SNR)}_{n} = \frac{\int_{0}^{\omega_n} E
\{{|\alpha_{n}(\omega)|}^2\} {|J_n(\frac{\omega}{c}R)|}^2 d\omega
}{\int_{0}^{\omega_n} \sigma_0 ^2 d\omega}.
\end{eqnarray}
Note that we consider white noise and is independent of frequency.
Hence, using \eqref{k}, \eqref{snr} can be rewritten as
\begin{eqnarray} \label{snr2}
{(SNR)}_{n} \leq {(SNR)}_{\max} \frac{{(R/c)}^{2n}}{\omega_n
{2}^{2n} {[\Gamma(n+1)]}^2} \int_{0}^{\omega_n} {\omega}^{2n}
d\omega.
\end{eqnarray}
This result is obtained based on the fact that for large order $n$,
the Bessel functions can be approximated as \cite[eqn
9.1.7]{AbramowitzStegun}
\begin{eqnarray}\label{eq:bapprox}
J_n(z)\sim {(\frac{1}{2}z)}^n/ \Gamma(n+1), & \mbox{$n\geq 0$}
\end{eqnarray}
where $\Gamma(\cdot)$ is the Gamma function. Now, we use the
Stirling lower bound on the Gamma functions, $\Gamma(n+1)>
\sqrt{2\pi n} n^n e^{-n}$, to write \eqref{snr2} as
\begin{eqnarray} \label{snr3}
 {(SNR)}_{n} &<& {(SNR)}_{\max} \frac{1}{2\pi n (2n+1)} {\left(\frac{e \omega_n
R/c}{2n}\right)}^{2n} \nonumber \\
&<& {(SNR)}_{\max} e^{-(2n-2\pi e F_n R/c)}
\end{eqnarray}
since $\beta = 1/(2\pi n (2n+1))< 1$ and using the exponential
inequality, $(1+ x/n)^n \leq e^x$ for $n \neq 0$.

Note that for the $n^{th}$ order, the $(SNR)_n$ must be larger than
the threshold $\gamma$, in effect,
\begin{equation}\label{eq:detec}
{(SNR)}_{\max} e^{-(2n-2\pi e F_n R/c)} \geq \gamma
\end{equation}
which results in \eqref{cf}.
This means that for order $n$, signals below frequency $F_n$ are not
detectable since \eqref{eq:detec} will not be satisfied. Observe
that for any particular order $|n|(>0)$, if $F_n>F_0-W$, the
effective bandwidth of that order is $F_0+W-F_n$. In addition, if
$F_n>F_0+W$, the effective bandwidth of this order and orders above
this is zero and we can truncate the infinite series in
\eqref{field1} to $|n|< N_u$. These arguments can be written
mathematically as \eqref{BW}.
\end{IEEEproof}

\section{Degrees of Freedom of $2D$ Multipath Fields}

In this section, we derive an expression to estimate the degrees of
freedom available in a wideband multipath field observable over
finite time and space windows. Note that so far we showed that any
wireless multipath wavefield band limited to $[F_0-W,F_0+W]$ and
observed within a $2D$ disk region of radius $R$ over a finite time
interval $[0,T]$ can be represented as a series of orthogonal basis
functions encoded in a finite number of orders $n$. A simple
observation based on this representation is that for each order $n$,
the observation time over space $T_{eff}$ is fixed \eqref{TS}.
Whereas, for each order $n$, the effective frequency bandwidth $W_n$
is different \eqref{BW}.

Note that the work of Shannon \cite{shannon49} provided a theorem to
determine the degrees of freedom available in a wideband channel
observed over a finite time interval for point to point
communications. We can think of Shannon's model as a wavefield
encoded in only one spatial order. Hence, following the classical
degrees of freedom result of {\em time-bandwidth product+ 1}, the
available degrees of freedom for each order is $W_nT_{eff}+1$. Therefore, we can evaluate the total degrees of
freedom available in our observable $2D$ multipath field as
\begin{eqnarray} \label{D}
D= \sum_{|n|< N_u} (W_nT_{eff}+1) = \sum_{|n|< N_u} \left[W_n\left( T+\frac{2R}{c} \right)+1\right]
\end{eqnarray}
where $W_n$ is given by \eqref{BW}. Note that the degrees of freedom
result in \eqref{D} does not agree with the well established result
of evaluating degrees of freedom of spatially diverse wideband
wavefields as a product of {\em space-time-bandwidth}
\cite{PoonBrodersenTse,Franceschetti2012}. However, in the
propagation of waves even though space, time and frequency are
separate entities, in spatially diverse wideband wavefields space
and time as well as space and frequency are strongly coupled, the
results of \cite{PoonBrodersenTse,Franceschetti2012} fail to show
those coupling relationships. On the contrary, our derived result
clearly indicates the coupling relationships between space and time
as well as space and frequency.

\section{Conclusion}
\label{conclusion}

In this paper, we express any band limited wireless multipath
wavefield observed within a $2D$ disk region of finite radius over a
finite time interval as a series of orthogonal basis functions
encoded in a finite number of spatial orders. Our analysis shows
that (i) the effective observation time varies from given
observation time and is not spatial order dependent, and (ii) the
lower spatial orders can utilize the full frequency bandwidth,
whereas, the higher orders can utilize only a fraction of the given
bandwidth. These findings portray the strong coupling relations
between space and time as well as space and frequency. Thus, our
derived degrees of freedom result based on these findings clearly
indicates how the coupling relations impact the available degrees of
freedom of any $2D$ wideband multipath field observed over finite
time and space windows. We also show that the degrees of freedom is
affected by the acceptable SNR in each spatial order.


\bibliographystyle{IEEEbib}

\end{document}